\documentclass[prl,twocolumn,showpacs,superscriptaddress,preprintnumbers]{revtex4-2}
\usepackage{amsmath,amssymb,amsfonts,float,graphics,epsfig,epstopdf,color,verbatim,tabularx,bm,multirow,appendix}
\usepackage{tikz}
\usepackage{amsmath,amssymb,graphicx}
\usepackage{wasysym}
\usepackage[utf8]{inputenc}
\usepackage[T1]{fontenc}
\usepackage{xcolor}
\usepackage{dsfont}
\usepackage{tabularx}
\usepackage{array}

\newcommand{\tcell}[1]{\makebox[1.7cm][c]{#1}}
\usepackage{textcomp}
\usepackage{bm}

\usepackage{graphicx}%
\usepackage{xcolor}
\usepackage{dcolumn}
\usepackage{bm}
\usepackage{xspace}
\usepackage{time}
\usepackage{booktabs}
\usepackage{multirow}
\usepackage{makecell}

\pdfoutput=1
\usepackage{color}
\definecolor{LinkColor}{rgb}{0.256,0.439,0.588}
\usepackage{hyperref}
\hypersetup{
colorlinks=true,
citecolor=LinkColor,
linkcolor=LinkColor,
urlcolor=LinkColor
}

\def\bk{{\mathbf{k}}}

\def\bR{{\mathbf{R}}}
\def\br{{\mathbf{r}}}
\def\bq{{\mathbf{q}}}
\def\bQ{{\mathbf{Q}}}
\def\bG{{\mathbf{G}}}

\def\s{{\mathrm{s}}}

\newcommand{\newsect}[1]{\noindent \textit{\textcolor{blue}{#1.-}}}

\newcommand{\bra}[1]{\left \langle#1\right \rvert}
\newcommand{\ket}[1]{\left \lvert#1\right \rangle}

\usepackage{bibunits}

\begin{document}
\begin{bibunit}[apsrev4-2]
\title{Trion Excitations in Twisted Bilayer Graphene: A Quantum Monte Carlo Study}

\author{Shibo Shan}
\affiliation{Department of Physics and HK Institute of Quantum Science \& Technology, The University of Hong Kong, Pokfulam Road,  Hong Kong SAR, China}
\affiliation{State Key Laboratory of Optical Quantum Materials, The University of Hong Kong, Pokfulam Road,  Hong Kong SAR, China}

\author{Cheng Huang}
\affiliation{Department of Physics and HK Institute of Quantum Science \& Technology, The University of Hong Kong, Pokfulam Road,  Hong Kong SAR, China}
\affiliation{State Key Laboratory of Optical Quantum Materials, The University of Hong Kong, Pokfulam Road,  Hong Kong SAR, China}

\author{Patrick Ledwith}
\affiliation{Department of Physics, Massachusetts Institute of Technology, Cambridge, MA 02139, USA}

\author{Zi Yang Meng}
\email{zymeng@hku.hk}
\affiliation{Department of Physics and HK Institute of Quantum Science \& Technology, The University of Hong Kong, Pokfulam Road,  Hong Kong SAR, China}
\affiliation{State Key Laboratory of Optical Quantum Materials, The University of Hong Kong, Pokfulam Road,  Hong Kong SAR, China}

\date{March 2026}

\date{March 2026}

\begin{abstract}
Determining the nature of charge carriers is a fundamental goal in the study of strongly correlated electron systems. Here, we employ the continuous-field momentum-space quantum Monte Carlo method to reveal exotic ``Dirac trion'' excitations in the finite-temperature normal state of twisted bilayer graphene. While the ground state is a symmetry-breaking insulator with gapped ($\sim$ 20 meV) electron-like excitations, we show that a small temperature ($\sim$ 3 meV), well below the interaction scale, drives the system into a strongly fluctuating symmetric normal state. We demonstrate that this normal state hosts gapless excitations consisting of three-particle bound states, two electrons and one hole, that are exactly orthogonal to the higher-energy electrons at the zero-momentum gapless point. These Dirac trions have the remarkable property of being arbitrarily light despite being composed of heavy constituents, and their spectra can be easily tuned by varying the twist angle and interlayer hopping strength. Our unbiased quantum many-body computation sheds light on the Dirac trions in a projected correlated flat-band setting and opens the door for further investigation of many-body excitations in strongly correlated topological bands beyond Landau levels.

\end{abstract}

\maketitle

\newsect{Introduction} A central goal in condensed matter physics is characterizing the nature of collective excitations in strongly interacting electron systems. An understanding of charged excitations is especially important as they form the basis of doped phases. Moir\'e materials, such as twisted bilayer graphene\cite{EvaAndrei2010,bistritzerMoire2011,andrei2020graphene,nuckolls2024microscopic,ledwithreview2021} (TBG), introduce a new setting for strong correlations: narrow topological bands which, unlike Landau levels, can have inhomogeneous charge density in real space and inhomogeneous quantum geometry in momentum space. Thus far, significant progress has been made in showing that quantum Hall physics can be recovered in such bands~\cite{spanton2018observation,Ledwith2020,abouelkomsan2020particle,repellinChernBandsTwisted2019,wilhelm2021interplay,xie2021fractional,parkerfieldtuned,cai2023signatures,zeng2023thermodynamic,wu2019topological,Kaisun_FCI21,devakul2021magic,Valentin22_anomaloushallmetal,Cano_pressure_23,dong2023composite,goldman2023zero,park_observation_2023,lu_fractional_2024} despite these deformations from traditional Landau levels~\cite{kapitmueller,Ledwith2020,Wang2021,ledwith2022family,wang2022hierarchy,Dong2023Many,wang2023origin,ledwithVortexability2023,Liu_review_2023,huComposite2026}. On the other hand, TBG additionally hosts phenomena conventionally associated with Mott physics, such as unconventional superconductivity bordering a correlated insulator~\cite{caoCorrelated2018,caoSuperconductivity2018,Dean-Young,efetov,EfetovScreening,YoungScreening,NadjPergeSC,LiVafekscreening,ohEvidenceUnconventionalSuperconductivity2021,banerjee2025superfluid,tanaka2025superfluid,parkExperimental2026,hyunjinResoving2026,KwanKekule,nuckolls2023quantum,Kim2023imaging} as well as large, $O(k_B)$, magnetic entropy per site~\cite{rozen2021Entropic,saitoIsospinPomeranchukEffect2021,zhangHeavy2025}. Under the circumstances, one would like to ask: what new phenomena can emerge from this interplay between quantum Hall and Mott physics and, more broadly, topological bands far from the homogeneous Landau level limit?

Here we apply state-of-the-art continuous field momentum-space quantum Monte Carlo (QMC)~\cite{zhangMomentum2021,hofmannFermionic2022,zhangFermion2022,huangEvolution2024,huangAngle2025,huangStrain2026} to the interacting narrow bands of the Bistritzer-MacDonald (BM) model~\cite{tramblyLocalization2010,bistritzerMoire2011} of twisted bilayer graphene at charge neutrality. We focus on the exotic finite-temperature normal state, which emerges from the Kramers' intervalley coherent (KIVC) insulator ground state~\cite{bultinckGround2020,TBGIVGroundState,Vafek2020,ledwithreview2021,liaoCorrelation2021,liaoCorrelated2021,hofmannFermionic2022,panDynamical2022,TBGVI,Potasz,Tomo,Parker,stauberkivc,kong2025interacting} above a small temperature scale (much smaller than the interaction scale). This normal state has nearly decoupled flavor moments and, almost everywhere in the Brillouin zone, gapped charge. The $\Gamma$ point of the Brillouin zone, where the Berry curvature of the band is highly concentrated, however, hosts gapless charged excitations. This ``Mott semimetal'' spectrum\cite{ledwithNonlocal2024} is consistent with prior numerical~\cite{schindler2022trions,hofmannFermionic2022,panThermodynamic2023,huangEvolution2024,Haule2019,Calderon2020,Datta2023,Rai2024,calugaruObtainingSpectralFunction2025,hofmann2026montecarlostudiestwisted} and analytic~\cite{ledwithNonlocal2024,ledwithExotic2025,Hu2025,Lau2025,zhao2025ancillatheorytwistedbilayer,zhao2025mixed,vituri2026controlled,hu2026twisted,wei2026lifetime,nosov2026controlledexpansioncorrelatedelectrons} studies of the TBG normal state.

We show that the lowest-energy charged excitations of the Mott semimetal are ``Dirac trions:'' exotic three-particle bound states that emerge from the interplay between strong interactions and concentrated band topology~\cite{schindler2022trions,ledwithExotic2025,zhao2025ancillatheorytwistedbilayer,zhao2025mixed}. To do so, we compute the appropriate trion Green's function in QMC. As will be shown below, the trion operator contains 3 fermion operators; the trion Green's function is a six-fermion correlator whose direct momentum-space evaluation scales as \(O(N^4)\) (more than the QMC update itself, $O(N^3)$, with $N$ the system size). We overcome this prohibitive scaling through the use of a real-space representation of the topological flat bands, which simplifies the complexity to $O(N^2)$ (see below) and demonstrates a concrete and computationally useful application of recently developed real-space approaches to topological bands~\cite{zang2022realspacerepresentationtopological,schindler2022trions,huang2024coherent,gunawardanaOptimallyLocalizedSingleband2024,li2024constraints,xieChernBandsOptimally2024,okuma2024constructingvortexfunctionsbasis,ledwithNonlocal2024,ledwithExotic2025,cole2025reducedwannier,gerhard2026spatiallocalizerelectronsinsulators,becker2025fragile,becker2026wannier,verma2025localbasisinteractingtopological}. 

\begin{figure}[htp!]
\centering
\includegraphics[width=\columnwidth]{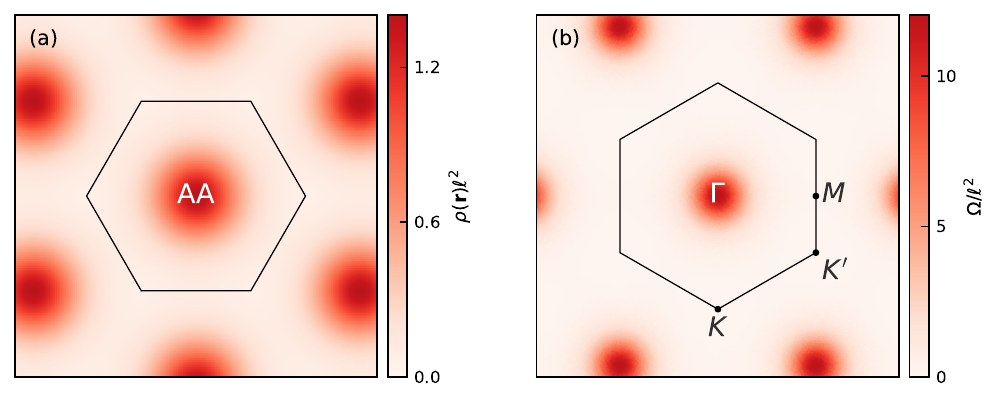}
\caption{
\textbf{Charge density and Berry curvature of the moir\'e flat bands.}
(a) Real-space charge density \(\rho(\mathbf r)\ell^2\) of the two flat bands, with
\(\ell=\sqrt{A_M/(2\pi)}\) and \(A_M\) the moir\'e unit-cell area. The black hexagon marks one moir\'e unit cell and the AA region is indicated.
(b) Momentum-space Berry curvature \(\Omega(\mathbf k)/\ell^2\) of the \(+1\) Chern band in the moir\'e Brillouin zone. 
The parameters are \(u_0=80\,{\rm meV}\), \(u_1=110\,{\rm meV}\), and \(\theta=1.08^\circ\).
}
\label{fig:fig1}
\end{figure}

We find that the trion spectrum has a zero-energy band touching at $\Gamma$, and verify that the $\Gamma$ point trion is exactly orthogonal to the electron as enforced by the band topology~\cite{schindler2022trions,ledwithExotic2025,zhao2025ancillatheorytwistedbilayer,zhao2025mixed}. In contrast, electrons at $\Gamma$ are split by the single particle dispersion. Exactly at the magic angle, this splitting vanishes and electron and trion are both gapless. We find that the trion near $\Gamma$ is very light, despite the fact that its constituent electrons and holes are heavy. We discuss how these features emerge from an analytic theory based on the limit of highly concentrated charge density (Fig. \ref{fig:fig1} (a)) and Berry curvature (Fig.~\ref{fig:fig1} (b))~\cite{ledwithNonlocal2024,ledwithExotic2025}. Finally, we comment on implications for twisted bilayer graphene experiments and future applications of the QMC method we develop.


\begin{figure*}[htp!]
    \centering
\includegraphics[width=0.99\textwidth]{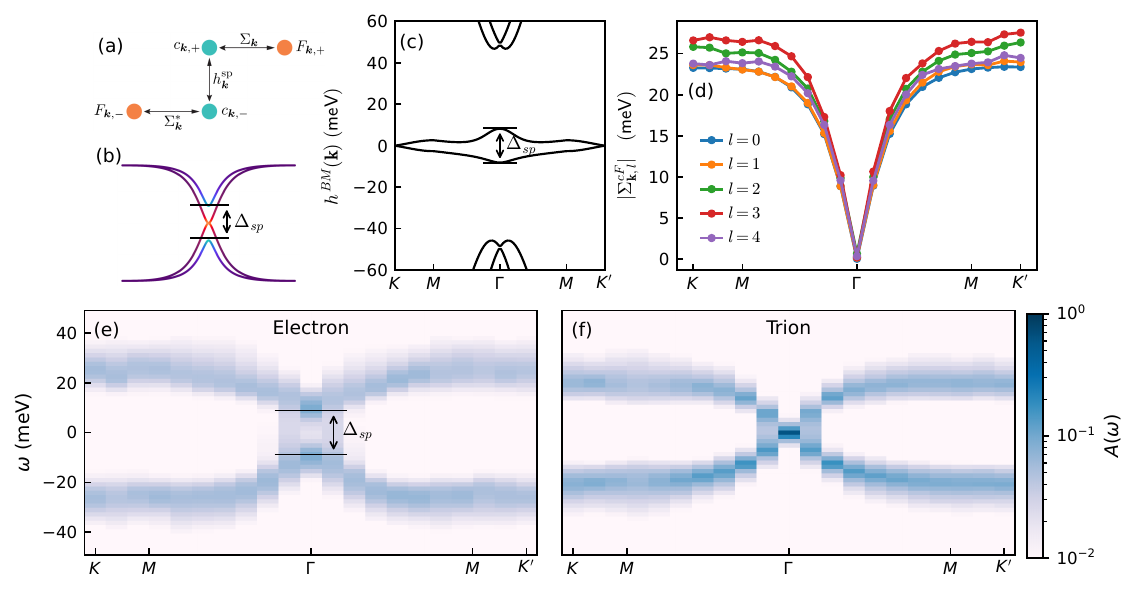}
    \caption{
    \textbf{Electron-trion hybridization and spectral function.}
    All data are obtained at \(u_0=80\,{\rm meV}\), \(T=3.5\,{\rm meV}\), and \(\theta=1.16^\circ\), away from magic angle.
    (a) Schematic analogy between the electron-trion problem and Bernal bilayer graphene, where one views the two electrons and two trions as the four sites in one unit cell of Bernal bilayer graphene. 
    (b)  Corresponding schematic band structure, with blue and orange denoting electron-like and trion-like bands, respectively.
    (c) BM model single-particle dispersion \(h^{BM}(\bk)\) along the high symmetry path \(K-M-\Gamma-M-K'\). For \(\theta=1.16^\circ\), the BM bands have a finite single-particle splitting near the \(\Gamma\) point.
    (d) Magnitude of the off-diagonal inverse-Green's-function component, \(|\Sigma^{cF}_{\bk,l}|\), along the same momentum path with $L = 15$. The legend labels \(l\) denote the fermionic Matsubara-frequency index, \(\omega_l=(2l+1)\pi T\). The data show that \(|\Sigma^{cF}_{\bk,l}|\propto|\bk|\) for small \(|\bk|\) around \(\Gamma\).
    (e,f) Electron and trion spectral functions with $L = 15$. The \(\Delta_{sp}\) markers in (b,c,e) indicate the same single-particle gap at \(\Gamma\).
    }
    \label{fig:fig2}
\end{figure*}

\newsect{Models and Observables} The kinetic part of our Hamiltonian is taken from the BM model~\cite{tramblyLocalization2010,bistritzerMoire2011,huangAngle2025} with various interlayer intrasublattice (AA) hopping $u_0$ and fixed interlayer intersublattice (AB) hopping $u_1=110$ meV. Since the gap separating the two flat bands from the remote bands is much larger than the interaction scale~\cite{xiaoInteracting2025}, we study the projected interaction Hamiltonian of the two bands
(See Refs.~\cite{huangAngle2025,huang2026strain} for the projection details)
\begin{equation}
H=\sum_{\bk ns\eta}h^{BM}_{n\eta}(\bk)c^\dagger_{\bk ns\eta} c_{\bk ns\eta}+\frac{1}{2\Omega}\sum_{\bQ}V(\bQ)\delta \rho_{-\bQ}\delta\rho_{\bQ},
\label{eq:eq1}
\end{equation}
where $\Omega$ is the sample area, $h^{BM}_{n\eta}(\bk)$ are the band energies~\cite{bistritzerMoire2011}, and $\bk$,\;$n=\pm1$,\;$s=\uparrow,\downarrow$,\;$\eta=\pm$ label moir\'e Brillouin zone (mBZ) momentum, band, spin, and valley respectively. $\delta \rho_\mathbf{Q}$ is the projected density operator at momentum $\mathbf{Q}$.
 We use the long-range single-gate screened Coulomb interaction $V(\bQ) =e^2\left(1-\mathrm{e}^{-|\bQ| d}\right)/\left(2 \epsilon|\bQ|\right)$, with permittivity $\epsilon=7\epsilon_0$ and gate distance $d/2=20$ nm~\cite{liuNematic2021}. 
 As shown in our previous work~\cite{panThermodynamic2023,panDynamical2022}, we can truncate $\bQ$ at a distance of order the moir\'e reciprocal-lattice scale. The validity of this projection has been explicitly examined in a complementary QMC study retaining remote BM bands~\cite{shanMomentumSpace2026}.
 
Following Ref.~\cite{ledwithExotic2025}, we define the trion operator as 
\begin{equation}
F_{\mathbf{k}\sigma s\eta}
= \frac{1}{\sqrt{N}}
\sum_{\mathbf{R}}
e^{-i\mathbf{k}\cdot\mathbf{R}}
\bigl\{ c_{\mathbf{R}\sigma s\eta},\, \delta n_{\mathbf{R}} \bigr\},
\end{equation}
where $\mathbf{R}$ labels moir\'e unit cells and \(\sigma=\pm\) labels the sublattice-polarized (Chern) basis of the two projected BM flat bands~\cite{bultinckGround2020,ledwithreview2021}. In this basis, \(c_{\bR\sigma s\eta}^{\dagger}\) creates an orbital polarized to the AA site $\bR$ and sublattice $\sigma$ (see Supplemental Material (SM)~\cite{suppl}), while \(\delta n_{\bR}\) is the corresponding local charge fluctuation relative to neutrality. Thus \(F_{\bR\sigma s\eta}=\{c_{\bR\sigma s\eta},\delta n_{\bR}\}\) is the analog of a local atomic-limit Hubbard trion~\cite{hubbardElectron1963}. At charge neutrality, it obeys \(\langle\{F_{\bk},c_\bk^\dagger\}\rangle=0\). We denote the BM single-particle Hamiltonian in this Chern basis by \(h_{\bk}^{sp}\); the explicit Fourier transforms, basis rotation, and matrix elements are given in the SM~\cite{suppl}. We pause to comment that the specific form of $F_{\bk}$ chosen here is convenient but not necessary; we only need to choose an operator that has appreciable overlap with the low-energy trion excitation.

To verify the existence of the low energy trion and understand its interplay with flat band electrons, we measure the correlation functions $\mathcal{G}_{AB}(\bk,i\omega_l) = \int_0^\beta d\tau e^{i\omega_l\tau}\mathcal{G}_{AB}(\bk,\tau)\), with \([\mathcal{G}_{AB}(\bk,\tau)]_{\sigma_1\sigma_2}=-\langle A_{\bk,\sigma_1}(\tau)B_{\bk,\sigma_2}^\dagger(0)\rangle\), \(A,B=c,F\), and the Matsubara frequency \(\omega_l=(2l+1)\pi T\). We will regard $\mathcal{G}$ as a (block) matrix in the basis \((c_{\bk+},c_{\bk-},F_{\bk+},F_{\bk-})_{s\eta}\), where $\pm$ label the sublattice $\sigma$.

We pause to discuss an analytic theory based on the concentration of charge density (Fig.~\ref{fig:fig1}(a)) and Berry curvature (Fig.~\ref{fig:fig1} (b)) of the TBG flat bands~\cite{ledwithNonlocal2024,ledwithExotic2025}. These features underscore strong deformation of the TBG bands away from the homogeneous Landau-level-like limit and drive the system's Mott-like physics. The theory is tractable due to a small parameter~\cite{ledwithNonlocal2024,ledwithExotic2025,nosov2026controlledexpansioncorrelatedelectrons,zhou2026quantummottsemimetalonedimensional}, $s^2 \ll 1$, where $s$ is the width of the Berry curvature distribution (see also recent generalizations to the topological heavy fermion model~\cite{Hu2025,vituri2026controlled,hu2026twisted,wei2026lifetime}) 
within the Mott regime $U \gg T \gg Us^2$ of mostly frozen charge and thermal flavor moments. 

In this limit the dynamics are controlled by an effective Hamiltonian
\begin{equation}
H_{\mathrm{eff}}(\bk)=
\begin{pmatrix}
0 & h_{\bk +-}^{sp} & \Sigma_{\bk} & 0 \\
h_{\bk -+}^{sp}& 0  & 0 & \Sigma_{\bk}^* \\
\Sigma_{\bk}^* & 0 & 0 & 0 \\
0 & \Sigma_\bk & 0 & 0 
\end{pmatrix} 
\label{eq:Green}
\end{equation}
such that $\mathcal{G}^{-1} =  i \omega_l - H_{\rm eff} + O(s^2)$. Refs.~\cite{zhao2025ancillatheorytwistedbilayer,zhao2025mixed} also proposed the trion excitation with the effective Hamiltonian in Eq.~\eqref{eq:Green} through an ancilla-based description~\cite{zhang2020pseudogap}. The $\Sigma_\bk$ terms arise from the Hubbard-like part of the projected Coulomb interactions; these on-site four-fermion terms tunnel electrons into three-particle trions and back~\cite{hubbardElectron1963,ledwithExotic2025}. Nonzero $h^{\rm{sp}}$ tunnels electrons between Chern/sublattice sectors~\cite{ledwithreview2021}.
 
The Hamiltonian Eq.~\eqref{eq:Green} maps onto that of Bernal bilayer graphene, where $\Sigma_\bk$ plays the role of the intralayer Dirac dispersion and $h^{\rm{sp}}$ acts as the interlayer tunneling (see Fig.~\ref{fig:fig2}(a)). Indeed, band topology enforces\cite{ledwithExotic2025} $\Sigma_\bk \propto k_x + i k_y$ at small $k$ (see SM~\cite{suppl}) such that for $h^{\rm sp} = 0$ we obtain two decoupled Dirac cones in the two ``layers'' of Fig.~\ref{fig:fig2}.  Nonzero $h^{\rm sp}$ couples the layers as in Bernal graphene, here generating gapped electron-like bands and a trion-like quadratic band touching. We therefore expect the lowest energy charged excitations to be trions that are orthogonal to the electron at $\bk = 0$. While controlled in models that introduce $s^2$ as a small parameter, this picture has not been established in the realistic BM model with long-range Coulomb interaction where unbiased quantum many-body numerics are required.
 

We confirm the above picture with QMC simulations. The electron-trion hybridization, $\Sigma^{cF}_{\bk,l} = - [\mathcal{G}^{-1}(\bk, i\omega_l)]^{cF}_{\sigma\sigma}$ is plotted in Fig.~\ref{fig:fig2}(d); it vanishes near $\Gamma$ and approaches the Mott gap in the rest of the mBZ. The electron spectral function, Fig.~\ref{fig:fig2}(e), features two peaks, split by the single-particle dispersion, whereas the trion spectral function, Fig.~\ref{fig:fig2}(f), is sharply peaked at zero energy, consistent with the Bernal-graphene-like Hamiltonian Eq.~\eqref{eq:Green}. These results verify that the low-energy degrees of freedom in the BM model are orthogonal light trions.

\newsect{Methods} We solve the Hamiltonian in Eq.~\eqref{eq:eq1} and measure the electron and trion correlation functions with the continuous field momentum-space QMC method~\cite{zhangMomentum2021,hofmannFermionic2022,huangAngle2025,huangStrain2026,hofmann2026montecarlostudiestwisted}. In the QMC simulation, we discretize the mBZ into an $N=L\times L$ momentum grid for $\bk$, with linear size $L=6,9,12,15$. 
Our method employs the Metropolis-adjusted Langevin algorithm (MALA)~\cite{huangAngle2025,fengScalable2026,wangHybrid2026,liaoNumerical2026} with global updates (see SM~\cite{suppl}). For the long-range Coulomb interaction in Eq.~\eqref{eq:eq1}, conventional local updates increase the cost from \(O(\beta N^3)\) to at least \(O(\beta N^4)\). In contrast, our global-update scheme simultaneously updates all auxiliary fields on each time slice and retains the \(O(\beta N^3)\) scaling. The resulting low autocorrelation also enables efficient measurement of the six-fermion trion Green's function.

In the QMC, we compute the two-point Green's functions \(G_{\bk_1n_1;\bk_2n_2}^{s\eta}(\tau_1,\tau_2)=\langle\mathcal{T}_\tau c_{\bk_1 n_1s\eta}(\tau_1)c^\dagger_{\bk_2 n_2 s\eta}(\tau_2)\rangle\) and other observables are then evaluated as combinations of these Green’s functions through Wick expansion for each auxiliary field configuration. However, a direct momentum-space measurement for \(G_{FF}\) is inefficient: since
\(F_{\mathbf{k}\sigma s\eta}\) contains 3 \(\bR\)-space fermion operators, the
correlator involves 6 fermion operators and 4 independent momentum
summations, giving an \(O(N^4)\) measurement cost for each \((\mathbf k,\tau)\), which is \(O(N)\) times more costly than the QMC update. Such a naive measurement would become the bottleneck of the simulation.

To reduce the measurement cost, we Fourier transform the Green's functions to
the real-space moir\'e basis,
\(G_{\mathbf R_1\sigma_1;\mathbf R_2\sigma_2}^{s\eta}
(\tau_1,\tau_2)\), using fast Fourier transforms with cost \(O(N^2\log N)\). The observable
\(G_{FF}\) is then evaluated through Wick
expansion in terms of real-space Green's functions, with only 2 real-space summations over $\mathbf R_1$ and $\mathbf R_2$ instead of the 4 momentum-space sums, reducing the subsequent
measurement complexity to \(O(N^2)\). Further details are given in SM~\cite{suppl}.

Spectral functions of both electron and trion are then obtained from \([G_{cc}(\bk,\tau)]_{\sigma\sigma}\) and \([G_{FF}(\bk,\tau)]_{\sigma\sigma}\) via the stochastic analytic continuation (SAC)~\cite{Sandvik1998Stochastic,beachIdentifying2004,shaoNearly2017}. Such a QMC+SAC scheme has been successfully applied in previous momentum-space QMC simulations to reveal the ground state and finite-temperature single-particle and collective excitation spectra of TBG~\cite{panDynamical2022,zhangSuperconductivity2021,huangEvolution2024,panThermodynamic2023,huangAngle2025}.

\begin{figure}[htp!]
\centering    \includegraphics[width=1\columnwidth]{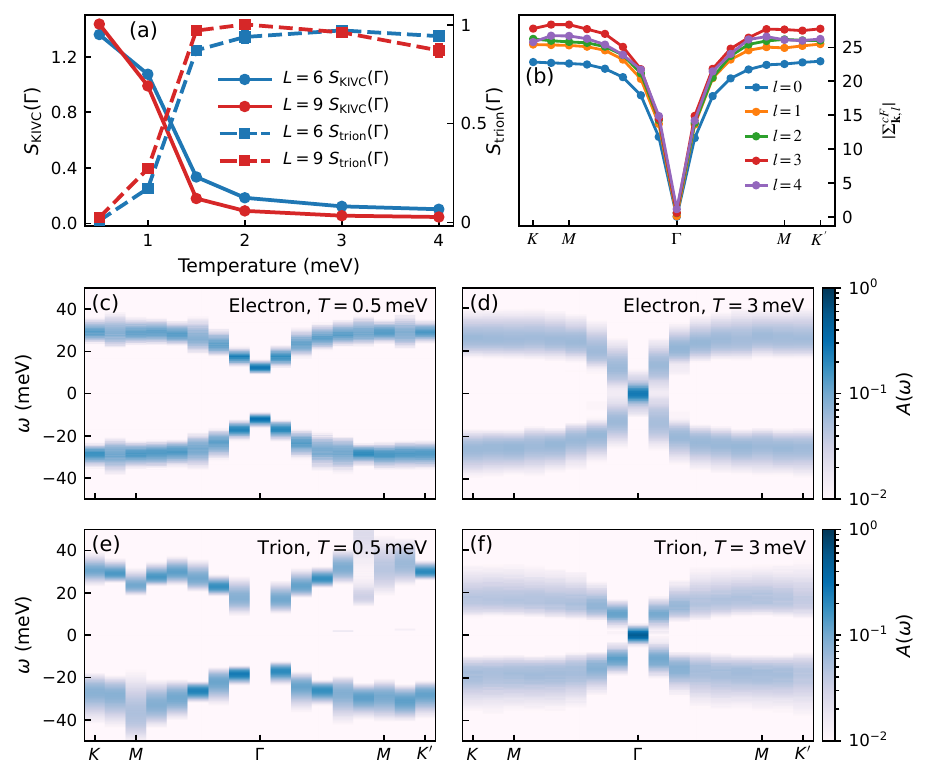}
\caption{\textbf{Temperature dependence of KIVC correlations, trion correlations, and spectral functions at the magic angle.}
    All data are obatained at \(u_0=80\,{\rm meV}\), and magic angle \(\theta=1.08^\circ\).
    (a) Temperature dependence of the KIVC structure factor \(S_{\mathrm{KIVC}}(\Gamma)\) and equal-time trion correlation function \(S_{\mathrm{trion}}(\Gamma)\) for system sizes \(L=6\) and \(L=9\).
    (b) Magnitude of the off-diagonal inverse Green's function component \(|\Sigma^{cF}_{\bk,l}|\) along \(K-M-\Gamma-M-K'\) at \(T=3\,{\rm meV}\). The data show that \(|\Sigma^{cF}_{\bk,l}|\propto|\bk|\) for small \(|\bk|\) around \(\Gamma\).
    (c,d) Electron spectral functions with \(L = 12\) at \(T=0.5\,{\rm meV}\) and \(T=3\,{\rm meV}\), respectively.
    (e,f) Corresponding trion spectral functions at the same temperatures.
    The melting of KIVC order with increasing temperature is accompanied by enhanced trion spectral weight concentrated at zero energy near \(\Gamma\).
    }
    \label{fig:fig3}
\end{figure}

\newsect{Results}
At $\theta=1.08^\circ$ and $\nu=0$, the ground state of Eq.~\eqref{eq:eq1} is known to be the  KIVC insulator~\cite{bultinckGround2020,liaoCorrelation2021,ledwithreview2021,liaoCorrelated2021,hofmannFermionic2022,panDynamical2022} with order parameter $O_{\mathrm{KIVC}}(\bq)
\equiv \sum_{\bk,s} c_{\bk+\bq,s}^{\dagger}\,\tau_xn_y\,c_{\bk,s}$. Here $c_{\bk,s}$ denotes the spinor in valley and band space, $\tau_x$ and $n_y$ are the Pauli matrices that act in the valley space and the two-flat-band space, respectively. In the finite-size QMC simulations, the valley U(1) symmetry is not broken, and we need to compute the correlation function of the KIVC order, $S_{\mathrm{KIVC}}(\bq)=\langle {O_{\mathrm{KIVC}}(\bq)}^\dagger O_{\mathrm{KIVC}}(-\bq)\rangle/N^2$, at its ordered wavevector $\bq=\Gamma$, to monitor the strength of the KIVC order. 

Fig.~\ref{fig:fig3} (a) shows how the KIVC order melts with increasing temperature. For $L=6,9$ systems, $S_{\text{KIVC}}$ vanishes quickly as $T>1$ meV. As the KIVC order breaks a continuous symmetry, it only has algebraic order at any $T>0$ in the thermodynamic limit. However, the electronic spectra of a fluctuating KIVC order with sufficiently large correlation length should be similar to the $T=0$ ordered state; hence, we observe the Mott semimetal to emerge when the flavor moment correlations are short-ranged or absent (such as $T>1$ meV here).

To detect the Mott semimetal state, we also monitor the temperature dependence of the static trion correlation function $S_{\text{trion}}(\Gamma)=\langle\{ F_{\Gamma\sigma s\eta}^\dagger, F_{\Gamma \sigma s \eta}\}\rangle$.
As shown in Fig.~\ref{fig:fig3} (a), trion spectral weight at \(\Gamma\) vanishes as $T\to 0$, where the KIVC order becomes long ranged, but becomes $\approx 1$ as the KIVC correlations become negligible in the Mott regime. 

The corresponding spectral evolution with temperature at $\theta=1.08^\circ$ is shown in Fig.~\ref{fig:fig3} (c-f). At \(T = 0.5\) meV, inside the KIVC regime, the electron spectrum is fully gapped, whereas the trion spectrum follows a similar dispersion away from \(\Gamma\) but carries nearly zero spectral weight at \(\Gamma\). At \(T=3\) meV, after the KIVC correlations have been strongly suppressed, the trion spectral weight is strongly enhanced at \(\Gamma\), and both the electron and trion spectra develop gapless Dirac cones centered at \(\Gamma\). The Dirac character is further supported by Fig.~\ref{fig:fig3} (b), which shows that the magnitude of electron-trion hybridization \(\Sigma^{cF}_{\bk,l}=-[\mathcal{G}^{-1}(\bk,i\omega_l)]_{\sigma\sigma
}^{cF}\) vanishes linearly on approaching \(\Gamma\) with only weak Matsubara-frequency dependence, \(|\Sigma^{cF}_{\bk,l}|\propto|\bk|\), consistent with the expected form \(\Sigma_{\bk}\sim k_x+ik_y\). Figure.~\ref{fig:fig3}(f) therefore provides the first manifestation of the Dirac trion in a realistic and unbiased quantum many-body setting.

\begin{figure}[htp!]
    \centering
    \includegraphics[width=\columnwidth]{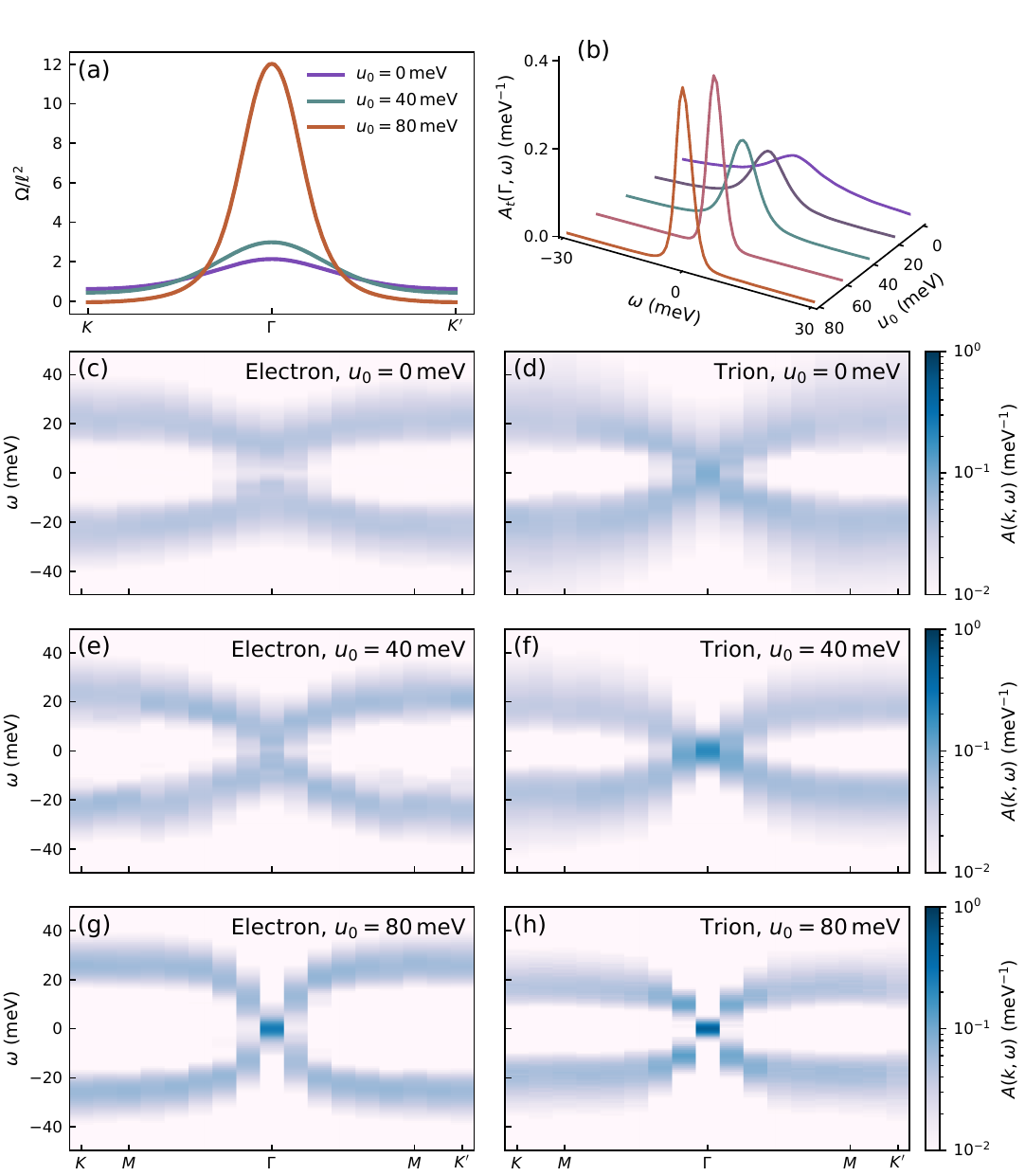}
    \caption{\textbf{Evolution of electron and trion spectra as a function of Berry curvature concentration.}
    (a) Berry-curvature line cuts along \(K-\Gamma-K'\) for \(u_0=0\), \(40\), and \(80\,{\rm meV}\) from BM flat bands at magic angle $\theta=1.08^\circ$.
    (b) Trion spectral function \(A_t(\Gamma,\omega)\) for \(u_0=0,20,40,60,\) and \(80\,{\rm meV}\). Increasing \(u_0\) produces a pronounced and sharply defined low-energy trion peak, accompanying the concentration of Berry curvature around \(\Gamma\).
    (c,d), (e,f), and (g,h) Electron and trion spectral functions for \(u_0=0\), \(40\), and \(80\,{\rm meV}\), respectively.
    The spectra are computed at \(L=12\), \(\theta=1.08^\circ\), \(T=3\,{\rm meV}\) along the high-symmetry path \(K-M-\Gamma-M-K'\). Increasing \(u_0\) concentrates the Berry curvature and enhances the low-energy trion feature near \(\Gamma\).
    }
    \label{fig:fig4}
\end{figure}
To further enhance the contrast between the electron spectrum $A_e(\mathbf{k},\omega)$ and the trion spectrum $A_t(\mathbf{k},\omega)$, we tune the twist angle a bit away from the magic angle from $\theta=1.08^\circ$ to $1.16^\circ$ at $T=3.5$ meV. 
At \(\theta=1.16^\circ\), the BM dispersion has a finite single-particle gap \(\Delta_{sp}\) at \(\Gamma\), as shown in Fig.~\ref{fig:fig2} (c). Consequently, the electron spectral function retains this gap in Fig.~\ref{fig:fig2} (e), in agreement with the analytic theory in Eq.~\eqref{eq:Green}. In contrast, the trion spectrum remains gapless and sharply peaked at \(\omega=0\) meV.

We further test the role of concentrated band topology by varying the interlayer AA hopping strength \(u_0\) at fixed \(\theta=1.08^\circ\) and \(T=3\) meV. As \(u_0\) is increased from the chiral limit (\(u_0=0\)) toward the realistic value (\(u_0=80\) meV), the Berry curvature becomes increasingly concentrated near \(\Gamma\), as seen from the line cuts in Fig.~\ref{fig:fig4}(a). In parallel, \(A_t(\Gamma,\omega)\) develops an increasingly pronounced peak at \(\omega=0\) in Fig.\ref{fig:fig4} (b). The full electron and trion spectra in Fig.\ref{fig:fig4} (c-h) show the same trend, with the electron and trion developing a more sharply defined low-energy Dirac-like feature near \(\Gamma\). This quantitative correspondence supports that Dirac trions arise from the combined effect of concentrated topology and strong Coulomb interactions.


\newsect{Discussion} 
In this work we systematically investigated the Dirac trion excitation in magic angle twisted bilayer graphene. Through sign problem free QMC, we showed that these trions are the elementary charged excitations in the Mott semimetal normal state. Furthermore, we verified that these low energy trions are exactly orthogonal to the electron. These bound states are highly dispersive, arbitrarily light near the magic angle, despite being composed of electrons and holes from the heavy parts of the band. These properties are unique amongst many-body excitations to our knowledge. Their emergence from the flat bands of twisted bilayer graphene, with concentrated charge density and Berry curvature, demonstrates that new exotic excitations can emerge from strongly correlated topological bands beyond Landau levels. We expect our computational techniques to be applicable and influential to future investigations in this timely direction.

Our computational demonstration of Dirac trions in the Bistritzer-Macdonald model suggests that these excitations are fundamental to realistic magic angle graphene. Indeed, a recent~\cite{xiaoInteracting2025} quantum twisting microscope (QTM)~\cite{inbarQuantum2023,weiqtm} study finds low electron spectral weight, and no electron Fermi surface, when doping integer filling states away from charge neutrality --- precisely where a light trion Fermi surface is expected theoretically. Future QTM experiments could provide indirect evidence for trions at neutrality by scanning low-strain samples\cite{kapferProgramming2023,tran2024quantitative} with slowly varying twist angle and observing the evolution of the electron spectral function.
More direct detection of trions may also be possible through phonon or photon assisted tunneling, as while the trion does not overlap with any single electron, it does overlap with electrons bound to particle-hole pairs.

\newsect{Acknowledgment}
We thank Kai Sun, Jörg Schmalian, Andrey V. Chubukov, Francisco Guinea and Yves H. Kwan for the inspiring discussion. P.L. thanks Eslam Khalaf, Ashvin Vishwanath, Junkai Dong, and Pavel Nosov for prior collaborations on closely related works. S.B.S, C.H. and Z.Y.M. acknowledge the support from the Research Grants Council (RGC) of Hong Kong Special Administrative Region (SAR) of China (Project Nos. AoE/P701/20, C7037-22GF, 17302223, 17301924, 17301725, AoE/P-604/25-R, 17302026), the ANR/RGC Joint Research Scheme sponsored by RGC of Hong Kong and French National Research Agency (Project No. A\_HKU703/22) and the State Key Laboratory of Optical Quantum Materials at HKU. P.L. is supported by the MIT Pappalardo Fellowship in Physics. We thank the HPC2021 system under the Information Technology Services at the University of Hong Kong~\cite{hpc2021}, as well as Beijing Paratera Tech Corp., Ltd~\cite{paratera} for providing HPC resources that have contributed to the research results reported in this paper.

\nocite{Grenander1994,Roberts1998,cooleyAlgorithm1965}
\putbib[bibtex]
\end{bibunit}
\makeatletter
\global\let\@FMN@list\@empty
\makeatother

\clearpage
\onecolumngrid

\begin{bibunit}[apsrev4-2]
\begin{center}
    \textbf{\large Supplemental Material for \\ 
    ``Trion Excitations in Twisted Bilayer Graphene: a Quantum Monte Carlo Study''}
\end{center}
\noindent{\centerline{Shibo Shan, Cheng Huang, Patrick Ledwith and Zi Yang Meng}}

\setcounter{section}{0}
\setcounter{secnumdepth}{3}
\setcounter{figure}{0}
\setcounter{equation}{0}
\setcounter{table}{0}
\renewcommand\thesection{S\arabic{section}}
\renewcommand\thefigure{S\arabic{figure}}
\renewcommand\theequation{S\arabic{equation}}
\renewcommand\thetable{S\arabic{table}}

\makeatletter
\renewcommand{\theequation}{S\arabic{equation}}
\renewcommand{\thefigure}{S\arabic{figure}}
\setcounter{secnumdepth}{3}
\makeatother

In this Supplemental Material, we provide technical details for the trion operator
construction and QMC measurement of the trion Green's function. In
Sec.~\ref{sec:SM1}, we construct the trion operator by introducing the AA-centered Wannier-like orbital used in the trion and explain its relation to the local charge channel of the
projected interaction. In Sec.~\ref{sec:SM2}, we describe the evaluation of
the trion Green's function in momentum-space QMC, emphasizing
the reduction of the computational cost in the QMC measurement from the direct momentum-space Wick
expansion to the real-space implementation. In Sec.~\ref{sec:SM3}, we
summarize the Metropolis-adjusted Langevin algorithm used for global
auxiliary-field updates. In Sec.~\ref{sec:analytictrions}, we provide an intuitive
explanation of the physical origin of Hubbard trions in TBG
and discuss the scope and limitations of the leading-order
analytical theory.

\section{BM model parameters}
The Fermi velocity is \(\frac{\hbar v_F}{\sqrt{3}a}=2377.45\) meV where \(a=1.42\)\AA\ is the carbon-carbon distance in graphene. The projected density operator $\delta \rho_\mathbf{Q}$, at momentum $\mathbf{Q}=\bq+\bG$, is
\(
\delta\rho_{\bQ}=\sum_{\bk, m,n,s,\eta}\lambda^{\eta}_{mn}\left(\bk,\bk+\bQ\right)\left(c^\dagger_{\bk ms\eta}c_{\bk+\bq ,ns\eta}-\frac{1}{2}\delta_{\bq,\mathbf0}\delta_{m,n}\right)
\), where \(\bq\) is in the mBZ and \(\bG\) is a reciprocal lattice vector. The form factor \(\lambda_{mn}^{\eta}(\bk,\bk+\bQ)=\langle u_{m\eta}(\bk)|u_{n\eta}(\bk+\bQ)\rangle\) is obtained from the BM Hamiltonian eigenstate \(|u_{n\eta}(\bk)\rangle\).

\section{construction of the Trion operator}
\label{sec:SM1}
As discussed in the main text, we define the trion operator as
\begin{equation}
F_{\mathbf{k}\sigma s\eta}
=
\frac{1}{\sqrt{N}}
\sum_{\mathbf R}
e^{-i\mathbf k\cdot\mathbf R}
\bigl\{
c_{\mathbf R\sigma s\eta},\delta n_{\mathbf R}
\bigr\},
\label{eq:eqS1}
\end{equation}
with
\begin{equation}
c_{\mathbf R\sigma s\eta}
=
\frac{1}{\sqrt N}
\sum_{\mathbf k\in{\rm mBZ}}
e^{i\mathbf k\cdot\mathbf R}
c_{\mathbf k\sigma s\eta},
\qquad
\delta n_{\mathbf R}
=
\sum_{\sigma,s,\eta}
c_{\mathbf R\sigma s\eta}^{\dagger}
c_{\mathbf R\sigma s\eta}
-4 .
\end{equation}
Here \(s\) and \(\eta\) denote spin and valley,  \(\sigma=\pm\)
labels the sublattice-polarized (Chern) basis within the two BM
flat bands~\cite{bultinckGround2020}. This basis is introduced so that the Fourier-transformed
operator \(c_{\mathbf R\sigma s\eta}^\dagger\) creates a projected orbital localized
near the AA region of the moir\'e unit cell. In this case,
\(\delta n_{\mathbf R}\) can be interpreted as the local charge fluctuation, making
\(\{c_{\mathbf R\sigma s\eta},\delta n_{\mathbf R}\}\) the natural
Hubbard-like local trion operator.

To construct this basis, we rotate the two BM energy bands at each
\(\mathbf k\) by diagonalizing the projected sublattice operator
\begin{equation}
\Gamma_{nn'}^\eta(\mathbf k)
=
\langle u_{n\eta}(\mathbf k)|
\sigma_z
|u_{n'\eta}(\mathbf k)\rangle ,
\end{equation}
and define
\begin{equation}
c_{\mathbf k\sigma s\eta}^{\dagger}
=
\sum_n
X_{\sigma n}^{\eta}(\mathbf k)
c_{\mathbf k n s\eta}^{\dagger},
\end{equation}
where \(|u_{n\eta}(\bk)\rangle\) is the eigenvector of BM Hamiltonian, \(\sigma_z\) acts on the microscopic graphene sublattice, and \(X_{\sigma n}^{\eta}(\mathbf k)\) are the eigenvectors of
\(\Gamma^\eta(\mathbf k)\). To make \(c_{\bR\sigma s\eta}^\dagger\) create an AA-centered orbital, we multiply
each eigenvector by a phase factor \(\chi_{\mathbf k\sigma}^{\eta}\), chosen
such that the top-layer \(A\)-sublattice component of the resulting Bloch
spinor created by \(c_{\bk \sigma s \eta}^\dagger\) is real and positive at the AA center. At the \(\Gamma\) point this top-layer \(A\)-sublattice component vanishes, \(c_{\Gamma\sigma s\eta}^\dagger\) cannot create a Bloch function concentrated at the AA center, so we exclude its contribution by setting \(X_{\sigma n}^{\eta}(\Gamma)=0\) in the trion operator. This gauge choice ensures that the
real-space wavefunction created by \(c_{\bR\sigma s\eta}^\dagger\)
\begin{equation}
w_{\mathbf R\sigma s\eta}^{\alpha l}(\mathbf r)
=
\langle \mathbf r,\alpha,l|
c_{\mathbf R\sigma s\eta}^{\dagger}
|0\rangle
\end{equation}
has a charge profile
\begin{equation}
\rho_{\mathbf R\sigma s\eta}(\mathbf r)
=
\sum_{\alpha,l}
\left|
w_{\mathbf R\sigma s\eta}^{\alpha l}(\mathbf r)
\right|^2
\end{equation}
\begin{figure}[t]
    \centering
    \includegraphics[width=0.85\textwidth]{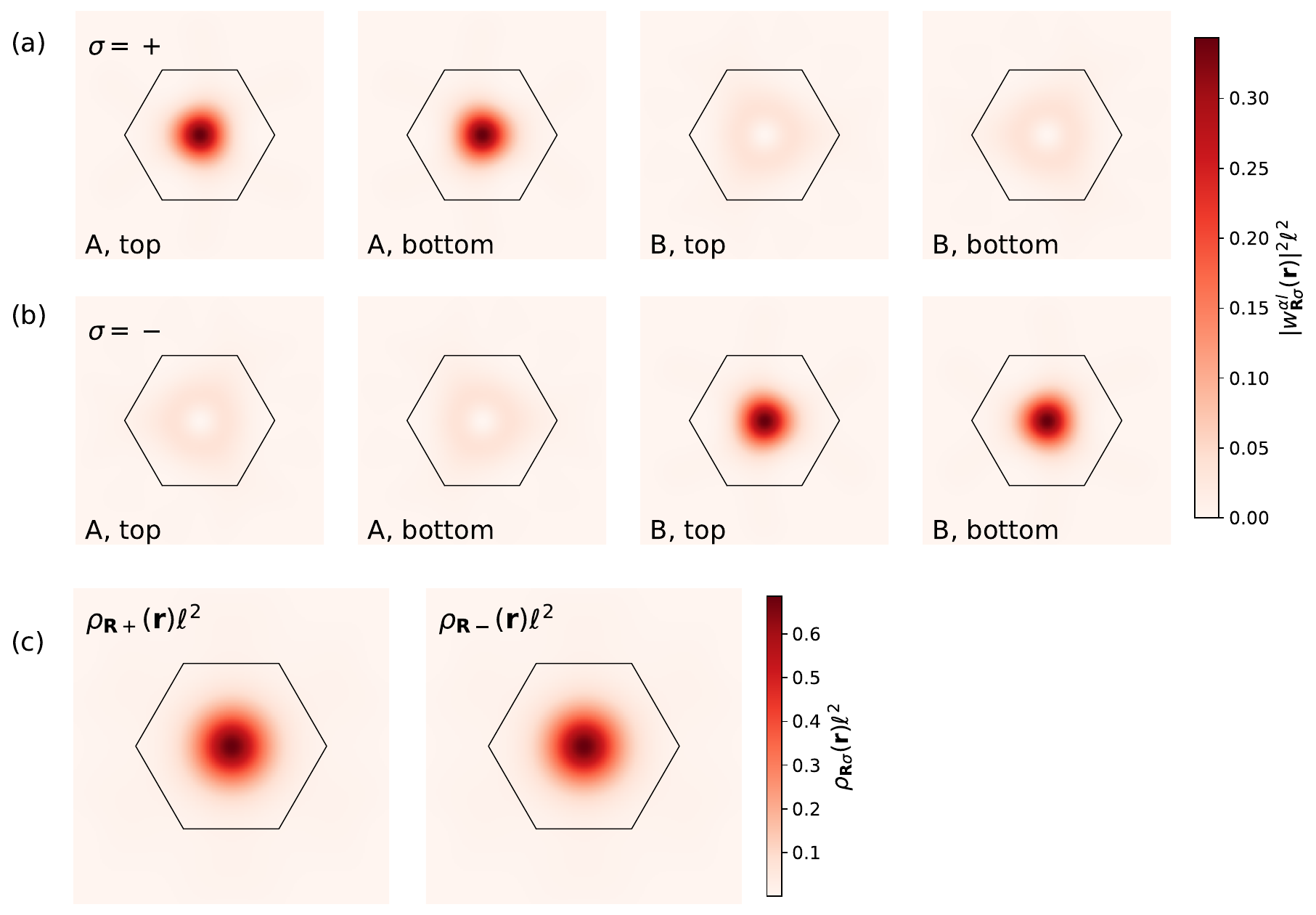}
    \caption{
    Real-space profiles of the projected Wannier orbitals in the
    sublattice-polarized basis for \(\theta=1.08^\circ, u_0=80\text{meV}\) and \(u_1=110\text{meV}\), at fixed spin \(s\) and valley \(\eta\).
    (a,b) Component densities
    \(|w_{\mathbf R\sigma s\eta}^{\alpha l}(\mathbf r)|^2\ell^2\)
    for \(\sigma=+\) and \(\sigma=-\), resolved by microscopic sublattice
    \(\alpha=A,B\) and layer \(l=\mathrm{top},\mathrm{bottom}\), with \(\ell=\sqrt{A_M/(2\pi)}\) and \(A_M\) the moir\'e unit-cell area.
    (c) Total charge profiles
    \(\rho_{\mathbf R\sigma s\eta}(\mathbf r)\ell^2=
    \sum_{\alpha,l}|w_{\mathbf R\sigma s\eta}^{\alpha l}(\mathbf r)|^2\ell^2\).
    The black hexagon marks one moir\'e Wigner-Seitz cell. The charge profiles
    are concentrated near the AA center, supporting the use of
    \(\delta n_{\mathbf R}\) as a local AA charge fluctuation in the trion
    operator.}
    \label{fig:wannier_function}
\end{figure}
concentrated near the AA center, as shown in Fig.~\ref{fig:wannier_function}, where \(\alpha=A,B\) and \(l = t,b\) are the microscopic sublattice and layer indices.

The above construction expresses the intrinsic AA concentration of the
flat-band wave functions in an AA-centered projected orbital basis. In this
basis, \(\delta n_{\mathbf R}\) measures the local AA charge fluctuation
associated with the leading Hubbard-like component of the projected
interaction. Thus \(\{c_{\mathbf R\sigma s\eta},\delta n_{\mathbf R}\}\)
is the projected analogue of the local Hubbard trion operator.

\section{Measurement of the Trion operator in QMC}
\label{sec:SM2}
In this section, we discuss the QMC measurement of the trion Green's function \(G_{FF}(\bk,\tau)\). For each auxiliary-field configuration \(\phi\), the fermionic problem is
quadratic. We define the corresponding time-ordered single-particle Green's function as
\begin{equation}
\begin{aligned}
G_{\bk_1n_1;\bk_2n_2}^{s\eta}(\tau_1,\tau_2)
={
\left\langle
\mathcal T_\tau
c_{\bk_1n_1s\eta}(\tau_1)c_{\bk_2n_2s\eta}^\dagger(\tau_2)
\right\rangle}_{\phi}
\label{eq:G_def_aux}
\end{aligned}
\end{equation}
which is directly computed in the QMC simulation~\cite{zhangMomentum2021,huangAngle2025}. \(\mathcal T_\tau\) is the imaginary-time-ordering operator and \(\langle\cdots\rangle_{\phi}\) denotes the fermionic value
evaluated for a specific auxiliary-field configuration \(\phi\). Note that Eq.~\eqref{eq:G_def_aux} adopts an auxiliary-field
contraction convention without the conventional overall minus sign used
for the physical fermionic Green's functions in the main text. For equal-time contractions, we use the one-sided convention
\(
G(\tau,\tau)
\equiv
\lim_{\epsilon\rightarrow0^+}
G(\tau+\epsilon,\tau).
\)
Then, the expectation value of observables \(\langle O\rangle\) can be obtained by first evaluating the Wick-expanded estimator \(\langle O\rangle_\phi\)  for
a fixed \(\phi\), and then averaging the estimator over the Monte Carlo
ensemble of auxiliary-field configurations as \(\langle O \rangle=\frac{1}{N_\phi}\sum_{\phi}\langle O \rangle_{\phi}\). For notational simplicity, in the following we suppress the auxiliary-field label \(\phi\) and leave the Monte Carlo average over auxiliary-field configurations implicit.

In $\mathbf{k}$ space basis we write Eq.~\eqref{eq:eqS1} with imaginary time $\tau$ dependence explicitly as 
\begin{eqnarray}
\begin{aligned}
F_{\bk\sigma s\eta}(\tau)&=\frac{1}{N}\sum_{\bk_1\bk_2\bk_3\sigma_1s_1\eta_1}\delta_{\bk+\bk_1-\bk_2-\bk_3=\bG}\{c_{\bk_3\sigma s\eta}(\tau),\,c_{\bk_1\sigma_1 s_1\eta_1}^\dagger(\tau) c_{\bk_2\sigma_1s_1\eta_1}(\tau)\}-8\,c_{\bk \sigma s\eta}(\tau)
\label{eq:eqS2}
\end{aligned}
\end{eqnarray}
We then compute the trion Green's function as \([G_{FF}(\bk, \tau)]_{\sigma_1\sigma_2}^{s\eta}=-\langle F_{\bk\sigma_1s\eta}(\tau)F_{\bk\sigma_2 s\eta}^\dagger(0)\rangle\). It is evaluated
by Wick expansion as products of sublattice-polarized basis Green's functions
\(G_{\bk_1\sigma_1;\bk_2\sigma_2}^{s\eta}(\tau_1,\tau_2)\), 
which is obtained from the
band-basis Green's function measured in QMC as
\begin{equation}
G_{\bk_1\sigma_1;\bk_2\sigma_2}^{s\eta}(\tau_1,\tau_2)
=
\sum_{n_1,n_2}
X_{\sigma_1 n_1}^{\eta*}(\bk_1)
X_{\sigma_2 n_2}^{\eta}(\bk_2)
G_{\bk_1 n_1;\bk_2 n_2}^{s\eta}(\tau_1,\tau_2),
\end{equation}
A direct momentum-space evaluation contains four independent internal
momentum sums in \(G_{FF}(\bk,\tau)\), leading to an \(O(N^4)\) cost for
each fixed \((\bk,\tau)\). This is even higher than the computational complexity of the global QMC update of $O(N^3)$ and becomes a bottleneck that severely limits the system sizes accessible and, therefore, limits the momentum resolution of the trion spectra. 

To overcome this problem in the trion Green's function measurement, we change the basis of the two-point Green's functions from momentum space into real space in the moir\'e scale $\mathbf{R}$ to reduce the computation complexity. From Eq.~\eqref{eq:eqS1} we have
\begin{eqnarray}
\begin{aligned}
F_{\bk\sigma s\eta}(\tau) &= \frac{1}{\sqrt{N}}\sum_{\bR\sigma^\prime s^\prime\eta^\prime}e^{-i\bk \cdot \bR}\Big(\{c_{\bR\sigma s\eta}(\tau),\ c^\dagger_{\bR\sigma^\prime s^\prime\eta^\prime}(\tau)c_{\bR\sigma^\prime s^\prime\eta^\prime}(\tau)\} -c_{\bR\sigma s\eta}(\tau)\Big)
\end{aligned}
\end{eqnarray}
Then the trion Green's function takes the form
\begin{equation}
[G_{FF}(\bk,\tau)]^{s\eta}_{\sigma_1\sigma_2}=-\langle F_{\bk\sigma_1 s\eta}(\tau)F_{\bk\sigma_2 s\eta}^\dagger(0)\rangle=\frac{1}{N}\sum_{\substack{\bR_1\sigma^\prime_1s^\prime_1\eta^\prime_1 \\
\bR_2\sigma_2^\prime s^\prime_2\eta^\prime_2}}e^{i\bk\cdot(\bR_2-\bR_1)}\langle O\rangle
\label{eq:realspace}
\end{equation}
where 
\begin{equation}
\langle O\rangle\equiv-\langle\big(\{c_{\bR_1\sigma_1\s\eta}(\tau),\,c^\dagger_{\bR_1\sigma_1^\prime s_1^\prime\eta^\prime_1}(\tau)c_{\bR_1\sigma_1^\prime s_1^\prime\eta_1^\prime}(\tau)\}-c_{\bR_1\sigma_1s\eta}(\tau)\big)\big(\{c_{\bR_2\sigma_2\s\eta}^\dagger(0),\,c^\dagger_{\bR_2\sigma_2^\prime s_2^\prime\eta^\prime_2}(0)c_{\bR_2\sigma_2^\prime s_2^\prime\eta_2^\prime}(0)\}-c_{\bR_2\sigma_2s\eta}^\dagger(0)\big)\rangle.
\label{eq:O}
\end{equation}
We then evaluate the term \(\langle O\rangle\) as combinations of \(G_{\bR_1\sigma_1;\bR_2\sigma_2}^{s\eta}(\tau_1,\tau_2)\) by Wick expansion, where 
\begin{equation}
G_{\bR_1\sigma_1;\bR_2\sigma_2}^{s\eta}(\tau_1,\tau_2)=\frac{1}{N}\sum_{\bk_1,\bk_2}e^{i(\bk_1\cdot\bR_1-\bk_2\cdot\bR_2)}G_{\bk_1\sigma_1;\bk_2\sigma_2}^{s\eta}(\tau_1,\tau_2).
\end{equation}
We evaluate this double Fourier transform using the fast Fourier transform (FFT) algorithm~\cite{cooleyAlgorithm1965}, which has a computational complexity of \(O(N^2\log(N))\). By evaluating the trion Green's function through Eq.~\eqref{eq:realspace}, we successfully reduce the 4 independent momentum loops to the 2 moir\'e unit cell \(\bR\) loops, with a computational complexity of \(O(N^2)\). The Wick expansion of the term \(\langle O\rangle\) is evaluated as \(\langle O \rangle=\sum_{i}a_iG_{i1}G_{i2}G_{i3}\), where \(a_i\), \(G_{i1}\), \(G_{i2}\), and \(G_{i3}\) are listed in the Table~\ref{tab:trion_green}. The convention \(X(\Gamma)=0\) in the trion operator construction changes the anticommutator of the real-space electron operators to
\begin{equation}
\{c_{\mathbf R a},c_{\mathbf R' b}^{\dagger}\}
=\left(\delta_{\mathbf R\mathbf R'}-\frac{1}{N}\right)\delta_{ab},
\qquad a,b=(\sigma,s,\eta).
\label{eq:modified_anticommutator}
\end{equation}
Therefore the explicit delta factors in the coefficient \(a_i\) of Table~\ref{tab:trion_green} should be read as
\begin{equation}
\delta_{x_2,y_2}\rightarrow
(1-1/N)\delta_{\sigma_2\sigma_2^\prime}\delta_{s s_2^\prime}\delta_{\eta\eta_2^\prime},
\qquad
\delta_{y_1,x_1}\rightarrow
(1-1/N)\delta_{\sigma_1\sigma_1^\prime}\delta_{s s_1^\prime}\delta_{\eta\eta_1^\prime}.
\label{eq:table_delta_replacement}
\end{equation}
The factor \(r=1-2/N\) in Table~\ref{tab:trion_green} also originates from this modified anticommutation relation.
By contrast, other observables such as \([G_{cc}(\bk,\tau)]_{\sigma_1\sigma_2}^{s\eta}\) or \([G_{cF}(\bk,\tau)]_{\sigma_1\sigma_2}^{s\eta}\) are still evaluated directly in momentum space by Wick expanding them in terms of
\(G_{\bk_1\sigma_1;\bk_2\sigma_2}^{s\eta}(\tau_1,\tau_2)\).

\begin{table}[htbp]
\centering
\renewcommand{\arraystretch}{1.4}
\begin{tabular}{c|c|c|c|c}
\hline
\tcell{\textbf{\(i\)}} 
& \tcell{\textbf{\(a_i\)}} 
& \tcell{\(G_{i1}\)} 
& \tcell{\(G_{i2}\)} 
& \tcell{\(G_{i3}\)} \\
\hline
\tcell{1} 
& \tcell{\(-r^2\)}
& \tcell{\(G_{x_1,x_2}(\tau,0)\)}
& \tcell{}
& \tcell{} \\
\hline
\tcell{2} 
& \tcell{\(r\delta_{x_2,y_2}\)}
& \tcell{\(G_{x_1,y_2}(\tau,0)\)}
& \tcell{}
& \tcell{} \\
\hline
\tcell{3} 
& \tcell{\(r\delta_{y_1,x_1}\)}
& \tcell{\(G_{y_1,x_2}(\tau,0)\)}
& \tcell{}
& \tcell{} \\
\hline
\tcell{4} 
& \tcell{\(-2r\)}
& \tcell{\(G_{x_1,y_1}(\tau,\tau)\)}
& \tcell{\(G_{y_1,x_2}(\tau,0)\)}
& \tcell{} \\
\hline
\tcell{5} 
& \tcell{\(2r\)}
& \tcell{\(G_{x_1,x_2}(\tau,0)\)}
& \tcell{\(G_{y_1,y_1}(\tau,\tau)\)}
& \tcell{} \\
\hline
\tcell{6} 
& \tcell{\(-2\delta_{x_2,y_2}\)}
& \tcell{\(G_{x_1,y_2}(\tau,0)\)}
& \tcell{\(G_{y_1,y_1}(\tau,\tau)\)}
& \tcell{} \\
\hline
\tcell{7} 
& \tcell{\(-\delta_{x_2,y_2}\delta_{y_1,x_1}\)}
& \tcell{\(G_{y_1,y_2}(\tau,0)\)}
& \tcell{}
& \tcell{} \\
\hline
\tcell{8} 
& \tcell{\(2\delta_{x_2,y_2}\)}
& \tcell{\(G_{x_1,y_1}(\tau,\tau)\)}
& \tcell{\(G_{y_1,y_2}(\tau,0)\)}
& \tcell{} \\
\hline
\tcell{9} 
& \tcell{\(-2r\)}
& \tcell{\(G_{x_1,y_2}(\tau,0)\)}
& \tcell{\(G_{y_2,x_2}(0,0)\)}
& \tcell{} \\
\hline
\tcell{10} 
& \tcell{\(4\)}
& \tcell{\(G_{x_1,y_2}(\tau,0)\)}
& \tcell{\(G_{y_1,y_1}(\tau,\tau)\)}
& \tcell{\(G_{y_2,x_2}(0,0)\)} \\
\hline
\tcell{11} 
& \tcell{\(2\delta_{y_1,x_1}\)}
& \tcell{\(G_{y_1,y_2}(\tau,0)\)}
& \tcell{\(G_{y_2,x_2}(0,0)\)}
& \tcell{} \\
\hline
\tcell{12} 
& \tcell{\(-4\)}
& \tcell{\(G_{x_1,y_1}(\tau,\tau)\)}
& \tcell{\(G_{y_1,y_2}(\tau,0)\)}
& \tcell{\(G_{y_2,x_2}(0,0)\)} \\
\hline
\tcell{13} 
& \tcell{\(-4\)}
& \tcell{\(G_{x_1,y_2}(\tau,0)\)}
& \tcell{\(G_{y_1,x_2}(\tau,0)\)}
& \tcell{\(G_{y_2,y_1}(0,\tau)\)} \\
\hline
\tcell{14} 
& \tcell{\(4\)}
& \tcell{\(G_{x_1,x_2}(\tau,0)\)}
& \tcell{\(G_{y_1,y_2}(\tau,0)\)}
& \tcell{\(G_{y_2,y_1}(0,\tau)\)} \\
\hline
\tcell{15} 
& \tcell{\(2r\)}
& \tcell{\(G_{x_1,x_2}(\tau,0)\)}
& \tcell{\(G_{y_2,y_2}(0,0)\)}
& \tcell{} \\
\hline
\tcell{16} 
& \tcell{\(-2\delta_{y_1,x_1}\)}
& \tcell{\(G_{y_1,x_2}(\tau,0)\)}
& \tcell{\(G_{y_2,y_2}(0,0)\)}
& \tcell{} \\
\hline
\tcell{17} 
& \tcell{\(4\)}
& \tcell{\(G_{x_1,y_1}(\tau,\tau)\)}
& \tcell{\(G_{y_1,x_2}(\tau,0)\)}
& \tcell{\(G_{y_2,y_2}(0,0)\)} \\
\hline
\tcell{18} 
& \tcell{\(-4\)}
& \tcell{\(G_{x_1,x_2}(\tau,0)\)}
& \tcell{\(G_{y_1,y_1}(\tau,\tau)\)}
& \tcell{\(G_{y_2,y_2}(0,0)\)} \\
\hline
\end{tabular}
\caption{Coefficients \(a_i\) and Green's functions \(G_{i1}\), \(G_{i2}\), and \(G_{i3}\) used in the Wick expansion of
\(\langle O \rangle\)
in Eq.~\eqref{eq:O}. Specifically, \(\langle O \rangle=\sum_i a_i G_{i1}G_{i2}G_{i3}\), with the \(i\)-th row representing the contribution 
\(a_iG_{i1}G_{i2}G_{i3}\). An empty entry is understood as the identity factor.
For compact notation, we define
\(x_1\equiv(\bR_1,\sigma_1,s,\eta)\),
\(y_1\equiv(\bR_1,\sigma_1^\prime,s_1^\prime,\eta_1^\prime)\),
\(x_2\equiv(\bR_2,\sigma_2,s,\eta)\), and
\(y_2\equiv(\bR_2,\sigma_2^\prime,s_2^\prime,\eta_2^\prime)\). For example, \(G_{x_1,y_2}(\tau,0)=\langle c_{\bR_1\sigma_1s\eta}(\tau)c_{\bR_2\sigma_2^\prime s_2^\prime \eta_2^\prime}^\dagger(0)\rangle=\delta _{ss_2^\prime}\delta_{\eta\eta_2^\prime}G_{\bR_1\sigma_1;\bR_2\sigma_2^\prime}^{s\eta}(\tau,0)\). Under the convention \(X(\Gamma)=0\), the explicit delta factors in \(a_i\) should be evaluated as \(
\delta_{x_2,y_2}=(1-1/N)\delta_{\sigma_2\sigma_2^\prime}\delta_{s s_2^\prime}\delta_{\eta\eta_2^\prime}\)
and
\(
\delta_{y_1,x_1}=(1-1/N)\delta_{\sigma_1\sigma_1^\prime}\delta_{s s_1^\prime}\delta_{\eta\eta_1^\prime}.
\) We also define \(r = 1-2/N\), which arises from the same convention.}
\label{tab:trion_green}
\end{table}



\section{Metropolis-Adjusted Langevin Sampling of the Auxiliary Fields}
\label{sec:SM3}
In this section, we describe the Metropolis-adjusted Langevin algorithm
(MALA) proposal used to sample the Hubbard--Stratonovich auxiliary fields
in our momentum-space determinant QMC simulation. After the Hubbard--Stratonovich decoupling of the density-density interaction, the auxiliary-field
configuration \(\phi\equiv\{\phi_{\tau,\bQ,1},\phi_{\tau,\bQ,2}\}\) is sampled from the probability distribution function~\cite{huangEvolution2024}
\begin{equation}
\pi(\phi)\propto e^{-\frac{1}{2}\sum_{\tau}\sum_{\bQ\in\text{half}}(\phi_{\tau,\bQ,1}^2+\phi_{\tau,\bQ,2}^2)}\text{Tr}_f(\prod_\tau e^{i\sum_{\bQ\in \text{half}}(-\phi_{\tau,\bQ,1}\sqrt{\alpha_2(\bQ)}A_\bQ+i\phi_{\tau,\bQ,2}\sqrt{\alpha_2(\bQ)}B_{\bQ})}e^{-\Delta\tau H_0})
\end{equation}
where \(\bQ = \bq+\bG\),  \(A_\bQ = \delta\rho_{-\bQ}+\delta\rho_{\bQ}\), \(B_{\bQ} = \delta\rho_{-\bQ}-\delta\rho_{\bQ}\), \(\alpha_2(\bQ)=\Delta\tau V(\bQ)/2\Omega\). The definition of \(\delta\rho_\bQ\) is given in the main text. The notation \(\sum_{\bQ\in\text{half}}\) means that only one momentum from each pair \(\pm\bQ\) is included. At charge neutrality, the
\(C_2P\) symmetry relates the two valleys and guarantees the absence of the
fermion sign problem~\cite{zhangMomentum2021}. The MALA proposal uses the local gradient of the logarithmic probability distribution. For a collective index \(i\equiv(\tau,\bQ,a)\), with \(a=1,2\), a new auxiliary-field configuration is proposed according to~\cite{Grenander1994}
\begin{equation}
\phi_i^{\prime}=\phi_i+\frac{\sigma^2}{2}\frac{\partial \ln(\pi(\phi))}{\partial\phi_i}+\sigma \xi_i 
\end{equation}
where each \(\xi_i\) is independently drawn from a standard normal distribution. The parameter \(\sigma\) controls the step size of the global
proposal. In practice, we tune \(\sigma\) such that the acceptance ratio is
close to the optimal high-dimensional MALA value \(0.574\)~\cite{Roberts1998}. The proposed configuration is accepted with probability
\begin{equation}
P(\phi\rightarrow\phi^\prime) = \text{min}(1,\frac{\pi(\phi^\prime)q(\phi|\phi^\prime)}{\pi(\phi)q(\phi^\prime|\phi)})
\end{equation}
where
\begin{equation}
 q(\phi^\prime|\phi)\propto \text{exp}\Big(-\frac{1}{2\sigma^2}\sum_{i}\Big(\phi_i^\prime-\phi_i-\frac{\sigma^2}{2}\frac{\partial\ln\pi(\phi)}{\partial\phi_i}\Big)^2\Big)
\end{equation}

We calculate \(\frac{\partial\ln(\pi(\phi))}{\partial\phi_i}\) as
\begin{equation}
\frac{\partial\ln\pi(\phi)}
{\partial\phi_{\tau,\bQ,1}}
=
-\phi_{\tau,\bQ,1}
+
4\sqrt{\alpha_2(\bQ)}
\,\mathrm{Im}
\left\{
\mathrm{Tr}
\left[
\left(I-G^{s\eta}(\tau,\tau)\right)
\mathcal A_{\bQ}^{s\eta}
\right]
\right\},
\label{eq:force_phi1}
\end{equation}
and
\begin{equation}
\frac{\partial\ln\pi(\phi)}
{\partial\phi_{\tau,\bQ,2}}
=
-\phi_{\tau,\bQ,2}
-
4\sqrt{\alpha_2(\bQ)}
\,\mathrm{Re}
\left\{
\mathrm{Tr}
\left[
\left(I-G^{s\eta}(\tau,\tau)\right)
\mathcal B_{\bQ}^{s\eta}
\right]
\right\}.
\label{eq:force_phi2}
\end{equation}
Here \(G^{s\eta}(\tau,\tau)\) is the equal-time single-particle Green's
function matrix in the \((\bk,m)\) basis,
\begin{equation}
[G^{s\eta}(\tau,\tau)]_{\bk_1m_1;\bk_2m_2}
=
\left\langle
c_{\bk_1m_1s\eta}(\tau)
c_{\bk_2m_2s\eta}^{\dagger}(\tau)
\right\rangle_{\phi}.
\end{equation}
The matrices \(\mathcal A_{\bQ}^{s\eta}\) and
\(\mathcal B_{\bQ}^{s\eta}\) are the single-particle matrix
representations of \(A_{\bQ}\) and \(B_{\bQ}\), with constant terms
omitted. Equivalently, their matrix elements are the coefficients of
\(c_{\bk_1m_1s\eta}^{\dagger}c_{\bk_2m_2s\eta}\) in the corresponding
bilinear operators.

\section{Analytic Theory of Hubbard Trions}

\label{sec:analytictrions}

In this appendix we discuss how trions emerge as the low energy excitations in twisted bilayer graphene. There are several references that provide full derivations, i.e. that start with a model that has $s^2$ as an explicit small parameter and study the model to leading order in $s^2$ thoroughly. We invite readers who want the full rigorous picture to read those refereces; in particular, Ref. \cite{ledwithNonlocal2024} introduces the model and provides a pedagogical introduction to the leading order $s^2 \ll 1$ diagrammatics, Ref. \cite{ledwithExotic2025} develops a more succinct algebraic approach and uses it to unveil the trion excitation (see also Refs. \cite{vituri2026controlled,wei2026lifetime,hu2026twisted,nosov2026controlledexpansioncorrelatedelectrons}, which develop the $s^2 \ll 1$ expansion to higher orders, and Refs. \cite{zhao2025ancillatheorytwistedbilayer,zhao2025mixed}, which derive the trion using an ancilla-based mean field theory.) 

Here we will take a higher-level approach to provide an intuitive picture for how the trions emerge. We refer readers who want to supplement this picture with a full derivation to one of the references above. It is useful to begin with how trions appear in the atomic limit of the Hubbard model:
\begin{equation}
  H_u = u \sum_{\bR} (\delta n_{\bR})^2.
  \label{eq:atomichubbard}
\end{equation}
Here $\delta n_{\bR} = n_\bR - 1$ is measured with respect to the charge neutral Mott insulator of one electron per site and $u=U/2$. The low temperature $T \ll U$ density matrix is 
\begin{equation}
  \rho =  \frac{1}{2}\left(\ket{\uparrow}\bra{\uparrow} + \ket{\downarrow}\bra{\downarrow}\right)
  \label{eq:densitymatrix}
\end{equation}
at each site, where $\uparrow$ and $\downarrow$ denote spin up and down electrons respectively. We will drop the site label $\bR$ below, as the analysis reduces to a single site problem.

The electron Green's function is
\begin{equation}
  G_{cc} = \frac{\frac{1}{2}}{\omega - u} + \frac{\frac{1}{2}}{\omega + u} = \frac{1}{\omega-\frac{u^2}{\omega}}.
  \label{eq:electrongreensfn}
\end{equation}
The first term corresponds to electron-like doublon excitations, whereas the second term corresponds to holon excitations. 
Even though the model is very simple, the Green's function is strongly interacting. There is no two-body effective Hamiltonian that can produce it; the self energy $u^2/\omega$ is strongly frequency dependent. This is not the case in an ordered state $\rho = \ket{\downarrow}\bra{\downarrow}$: then we would have $G^{-1} = \omega + Us_z$, where $U s_z$ is the effective Hamiltonian from the exchange interaction. However, as we will soon see, we can make the full Green's function take a non-interacting form if we introduce the trion $F_\sigma = \{c_\sigma, \delta n \}$. We will use this to develop an interpretation of on-site Hubbard dynamics as electron-trion tunneling via the interaction $u$. Remarkably, this interpretation appeared in Hubbard's original paper that introduced \eqref{eq:atomichubbard}. We will apply this interpretation to twisted bilayer graphene, where band topology leads to these trions becoming orthogonal to the electron and arbitrarily light near $\Gamma$.

The factors of $\frac{1}{2}$ in \eqref{eq:electrongreensfn} can be understood in two different ways. An electron $c^\dag_\uparrow$ has a $\frac{1}{2}$ chance to be Pauli blocked, and likewise for $c^\dag_\downarrow$, for the density matrix of fluctuating moments \eqref{eq:densitymatrix}. It will be also be instructive for us to think about acting with electrons in the $x$ basis, $c^\dag_{\rightarrow}$ and $c^\dag_{\leftarrow}$. Here \eqref{eq:electrongreensfn} arises from the $\frac{1}{2}$ overlap with the ``correct'' operators $c^\dag_{\downarrow} = \frac{1}{\sqrt{2}}(c^\dag_\rightarrow + c^\dag_\leftarrow)$, for the first term in \eqref{eq:densitymatrix}, and $c^\dag_\uparrow$ for the second term in \eqref{eq:densitymatrix}. 

In contrast, for an ordered state $\rho = \ket{\downarrow}\bra{\downarrow}$, electron excitations are spin up and cost energy $u$ and hole excitations are spin down and cost energy $-u$. This yields a simple Green's function $G^{-1} = \omega - H_F$ where $H_F = -u s_z$ is a Fock Hamiltonian that tunnels $c^\dag_{\rightarrow}$ into $c^\dag_{\leftarrow}$. For $\rho = \ket{\downarrow}\bra{\downarrow}$, the hybridization partner of $c^{\dag}_{\rightarrow}$ is $-c^\dag_{\leftarrow}$ instead. Another way to characterize this, for the $\ket{\downarrow}$ state for concreteness, is that the hybridization partner of $c^\dag_{\rightarrow}$ acts equivalently for electron-like excitations and oppositely for hole-like excitations:
\begin{equation}
  \begin{aligned}
  c^{\dag}_\leftarrow \ket{\downarrow} & = \frac{1}{\sqrt{2}}\ket{\uparrow \downarrow} = c^{\dag}_\rightarrow \ket{\downarrow}, \\
  c_{\leftarrow} \ket{\downarrow} & = -\frac{1}{\sqrt{2}}\ket{0} = c_{\rightarrow} \ket{\downarrow}.
  \label{eq:hybridizationacting}
  \end{aligned}
\end{equation}
This ensures that $\frac{1}{\sqrt{2}}(c^\dag_{\rightarrow} + c^\dag_{\leftarrow})\ket{\downarrow}$ and $\frac{1}{\sqrt{2}}(c^\dag_{\rightarrow} - c^\dag_{\leftarrow})\ket{\uparrow}$ have full overlap with the excited states $\ket{\uparrow \downarrow}$ and $\ket{0}$ respectively, in accordance with the $Z=1$ residue of the ordered state Green's function $G = 1/(\omega - Us_z)$. 
No one-body electron operator can function as the hybridization partner of $c_{\rightarrow}$ for the $\ket{\downarrow}$ and $\ket{\uparrow}$ states simultaneously. As a result, the symmetric Green's function \eqref{eq:electrongreensfn} cannot be described by an effective non-interacting Hamiltonian for electrons.

Now, consider the trion $F^\dag_{\rightarrow} = \{ c^\dag_{\rightarrow}, \delta n \}$. We can see that it \emph{always} acts as the hybridization partner of $c^\dag_{\rightarrow}$.
This is because, by construction, it acts like $c^\dag_{\rightarrow}$ for electron-like excitations, where $\delta n$ leads to a positive sign, and like $-c_{\rightarrow}$ for hole-like excitations, where $\delta n$ leads to a negative sign. For the state $\ket{\downarrow}$, the trion is equivalent to $c_{\leftarrow}$, whereas for the state $\ket{\uparrow}$ it is equivalent to $-c_{\leftarrow}$. As a result, the operator $\frac{1}{\sqrt{2}}(c^\dag_\sigma + F^{\dag}_\sigma)$, for any spin $\sigma$, maps the symmetric density matrix \eqref{eq:densitymatrix} to the full-normalization filled state $\ket{\uparrow \downarrow} \bra{\uparrow \downarrow}$ and likewise for $\frac{1}{\sqrt{2}}(c_\sigma - F_\sigma)$ and the empty state. The full Green's function in the electron-trion space is
\begin{equation}
  \mathcal{G} = \frac{1}{\omega - H_{\rm eff}}, \quad H_{\rm eff} = u \mu_x
  \label{eq:hubbardtrioneffective}
\end{equation}
where $\mu_x$ is the $x$ Pauli matrix in the electron-trion space. This corresponds to electrons tunneling into trions and vice versa with amplitude $u$. This is not a surprise: the Hubbard Hamiltonian is four body, so it naturally tuennels a one-body excitation into a three-body one. Indeed, $[H_u, c^\dag_\sigma] = u F^\dag_\sigma$, and $[H_u, F^\dag_\sigma] = u (2\delta n - 1)^2 c^\dag_\sigma$, which is equivalent to $u c^\dag_\sigma$ for single particle excitations on top of the Mott insulator.

\subsection{Origin of Trions in TBG}

While the above picture is in many ways elegant and theoretically appealing for the atomic Hubbard Hamiltonian, it makes some sense that it has not seen widespread use: the electrons and trions are strongly hybridized, such that the holon and doublon quasiparticles have $\frac{1}{2}$ overlap with the electron operators. They are essentially a specific form of dressed electron, and the electron operator suffices to create the excitation even if it doesn't have perfect overlap with it. 

In twisted bilayer graphene, however, the electron-trion hybridization vanishes at $\Gamma$, and the trions become sepearate, orthogonal, excitations. Thus, if we want to describe the low energy charged excitations, we must invoke them. We will discuss an intuitive picture for this in both momentum space and real space. 

We begin with momentum space. We note that the bulk of the Brillouin Zone in TBG, other than an $O(s^2)$ neighborhood of the $\Gamma$ point, is described by plane waves of localized states at the $AA$ centers of the moir\'e unit cells (with eight flavors corresponding to spin$\times$valley$\times$sublattice). These electrons undergo Hubbard dynamics, hybridizing with the corresponding trions $F_{\bk}$ and opening a Mott gap of $2u$. For $\bk = \Gamma$, however, the electrons have a distinct orbital character, with additional orbital angular momentum about the AA center: $\psi_{\Gamma,\gamma}(\br\approx 0) \propto x\pm iy$, where $\gamma = \pm$ labels the Chern sector\cite{ledwithreview2021,ledwithNonlocal2024} ($\gamma = \sigma \tau$, where $\sigma = \pm$ is sublattice and $\tau = \pm$ is the graphene valley). Their amplitude at the AA site, where the Hubbard-like interactions are concentrated, vanishes. This additional angular momentum can be seen as a consequence of band topology; the Chern number is the sum of the angular momenta at each high symmetry point $C = \ell_K + \ell_{K'} + \ell_\Gamma$. Relatedly, the sum $\sum_\bR e^{-i \bk \cdot \bR} c_\bR$ vanishes at $\bk = \Gamma$ if $c_\bR$ are chosen as exponentially localized coherent states, and is ill-defined at $\bk = \Gamma$ if they are chosen to be Wannier states with power law tails\cite{li2024constraints}; otherwise the full band could be Wannierized with the same angular momentum at each high symmetry point.

The trion $F^\dag_{\Gamma}$, however, is still a plane wave of localized states with no spatial angular momentum about the AA center. It is localized at the AA sites for all $\bk$. To see this, consider the trion operator definition
\begin{equation}
  F_\bk = \frac{1}{\sqrt{N}}\sum_\bR e^{-i \bk \cdot \bR} \{c_\bR, \delta n_\bR \}
  \label{eq:trionTBBsupp}
\end{equation}
The sum over $\bR$ does not vanish at $\bk = 0$ because of the $\delta n_\bR = \pm$ factors, which can appear before or after $c_\bR$ and contribute different signs from site to site. The trion is thus a plane wave of localized AA states for all $\bk$, unlike the electron. It also has trivial spatial angular momentum: $c_\bR \to c_{C_3 \bR}$ maps $F_\bk \to F_{C_3 \bk}$. One can also arrive at this conclusion by expanding out $c_{\bR}$ and $\delta n_\bR$ to arrive at a momentum space form, schematically $\sum_{\bk_1, \bk_2} c_{\bk_1}c^\dag_{-\bk-\bk_1-\bk_2} c_{\bk_2}$. Since $\bk_1$ and $\bk_2$ are summed over, the constituent electrons and holes in the trion generically have momenta far away from $\Gamma$ such that they are AA-localized.

Thus, while the electron and trion are both Hubbard like away from the $\Gamma$ point region, near $\Gamma$ they have distinct orbital character and angular momentum due to the electronic band topology. Their hybridization accordingly also has angular momentum around $\bk = \Gamma$. While away from $\Gamma$ the electron-trion tunneling is simply $\frac{U}{2}$, in a smooth gauge choice for $c_\bk$ near $\Gamma$ the tunneling must instead take the form $\Sigma_\bk = \lambda_\bk \frac{U}{2}$, where $\lambda_\bk \propto k_x + i \gamma k_y$ near $k = \Gamma = 0$. This precisely reproduces the electron-trion terms in the effective Hamiltonian \eqref{eq:hubbardtrioneffective}. 

Furthermore, we can understand the orthogonality of the electron and trion at $\bk = 0$ from their distinct orbital angular momentum. There is also a direct topological argument for this orthogonality. In a smooth gauge, the phase of $c_\bk$ winds by $2\pi C$ around the boundary of the Brillouin Zone, where $C$ is the Chern number. The trion operator, however, is smooth and periodic in $\bk$: its phase does not wind around the boundary of the Brillouin Zone. As a result, the overlap $\zeta_{\bk} = \langle c_\bk F^\dag_\bk \rangle$ winds by $2\pi$ around the boundary of the Brillouin zone. This is only possible if $\zeta_\bk$ has zeros, such that the electron and trion are orthogonal somewhere in the Brillouin Zone. This point is pinned to $\Gamma$ by $C_{3z}$ symmetry.

\subsubsection{Real space picture}

The above conclusion can also be understood in a real space Wannier basis. The Wannier states $\psi_{\bR}(\br)$ have almost all their weight, $1-O(s^2)$ at the AA site $\bR$. However, they also have a power law tail at long distances which leads to wavefunctions at different sites overlapping. The interacting Hamiltonian in the Wannier basis has the form 
\begin{equation}
\begin{aligned}
  H_{\rm int} & = \frac{1}{2}\int d^2 \br d^2 \br' V(\br - \br') \delta \rho_\br \delta \rho_{\br'}, \\
  \delta \rho_\br & = \sum_{\bR_1 ,\bR_2} \overline{\psi_{\bR_1 }(\br)} \psi_{\bR_2 }(\br)(c^\dag_{\bR_1 } c_{\bR_2 } - \tfrac{1}{2}\delta_{\bR_1,\bR_2}),
\end{aligned}
\label{eq:fullinteractingH}
\end{equation}
where $V(\br)$ is the (screened) Coulomb interaction and we suppress flavor indices for concision.
The Hamiltonian \eqref{eq:fullinteractingH} contains contributions from Wannier states at different sites: the tail of one or more states can interact with the dominant peak of the other states by directly overlapping in real space. For example, the Hubbard part of the interaction, where $\br,\br'$ are in the same concentrated charge density peak at a site $\bR$, includes the contributions
\begin{equation}
  \frac{U}{2} (\delta n_\bR)^2 + \frac{1}{2}\sum_{\bR' \neq \bR}  \mathcal{T}(\bR - \bR') \{ \delta n_\bR ,c^\dag_\bR \} c_{\bR'} + h.c. + \cdots,
  \label{eq:overlap}
\end{equation}
where
\begin{equation}
\begin{aligned}
  U & = \mathcal{T}(0) = \int_{\br,\br'\approx \bR} V(\br-\br') |\psi_\bR(\br)|^2 |\psi_\bR(\br')|^2 \\
  \mathcal{T}(\bR - \bR') & = \int_{\br,\br' \approx \bR} V(\br-\br') |\psi_\bR(\br)|^2 \overline{\psi_\bR(\br')} \psi_{\bR'}(\br')
  \end{aligned}
  \label{eq:coefficients}
\end{equation}
respectively correspond to the conventional Hubbard term and a term where one tail interacts with three on-site peaks. The latter can be identified as an electron trion hopping. 

The real-space properties of $\mathcal{T}$ reveal how we can use small $s^2$ to simplify the a priori extremely complex full Hamiltonian \eqref{eq:fullinteractingH}. For $\bR \neq \bR'$, $\mathcal{T}$ is $O(s^2)$ as there is little weight in the tails of $\psi$. However, the tails originate from a concentrated region of momentum space of width $s$; this small momentum space scale leads to a large characteristic correlated hopping range of $s^{-1}$ (beyond this scale the power law decay sets in\cite{ledwithNonlocal2024}). Thus, $\mathcal{T}$ leads to a low probability of hopping an electron at $\bR'$ to a far away trion at $\bR$ (and similarly for electron hopping via the single particle dispersion $h^{\rm sp}$). The probability of probability of returning to the same site is further diminished by $s^2$, as it is choosing a single site out of the $\sim s^{-2}$ available sites within a range of $s^{-1}$. This latter property is the key simplification, as hopping trajectories that do not self intersect can be summed into a geometric series\cite{nosov2026controlledexpansioncorrelatedelectrons}; schematically $\mathcal{G}_0 + \mathcal{G}_0 \mathcal{T} \mathcal{G}_0 + \cdots$, where $\mathcal{G}_0$ is the amplitude to stay at the same site (the atomic limit Green's function). The corresponding geometric series is then the full electron-trion Green's function to leading order in $s^2$:
\begin{equation}
    \mathcal{G}^{-1}_\bk = \omega - \frac{U}{2} \mu_x - h^{\rm sp}_\bk - \mathcal{T}(\bk)|_{\bR \neq \bR'} = \omega - \frac{U}{2} \mathcal{T}(\bk) - h^{\rm sp}_\bk.
  \label{eq:electrontriongreensupp}
\end{equation}
Here we initially exclude the $\bR = \bR'$ term in the Fourier transform of $\mathcal{T}$ in accordance with \eqref{eq:overlap}. But this is simply the on-site Hubbard electron trion hopping, such that we arrive at the second line of \eqref{eq:electrontriongreensupp} with
\begin{equation}
  \mathcal{T}(\bk) = \int_{\br,\br' \in \bR} V(\br-\br') |\psi_\bR(\br)|^2 \overline{\psi_\bR(\br')} \psi_{\bk}(\br').
  \label{eq:tofksupp}
\end{equation}
As we discussed above, at $\bk = 0$ the wavefunction $\psi_\bk(\br)$ has angular momentum about the AA centers. As a result, $T(\bk) \approx k_x \pm i k_y$, where $\pm$ labels the Chern sector, near $\bk = 0$. This matches the conclusions and effective Hamiltonian we quoted in the main text upon identifying $\frac{U}{2} \mathcal{T}(\bk) = \Sigma(\bk)$ in the positive Chern sector and $\overline{\Sigma(\bk)}$ in the negative Chern sector. 

We pause to comment that we have only selected the terms in \eqref{eq:fullinteractingH} that correspond to electron trion hopping. These terms turn out to be the only ones that contribute to the charge neutrality Green's function at leading order in $s^2$. There are other terms that also involve only two distinct sites and can form arbitrarily long chains; they lead to mean-field-like Hartree and Fock dispersions that vanish for the symmetric state at charge neutrality\cite{ledwithExotic2025,nosov2026controlledexpansioncorrelatedelectrons}. Other terms enter the Green's function at higher order in $s^2$.

We close by discussing some limitations of the above approach, as well as the fate of trions in the low temperature state with long-range KIVC correlations. First, the expansion is in a small parameter $s^2 \ll 1$. While its conclusions are likely correct for sufficiently small $s^2$, and in twisted bilayer graphene $s^2 \approx 0.06$, there are also eight flavors of electrons. For some contributions the relevant small parameter is $Ns^2 \approx 0.48$: considerably larger\cite{vituri2026controlled}. Thus, predictions of the leading order theory should be verified with numerics, such as the sign problem free QMC done in the main text. Second, the predictions from the leading order small $s^2$ theory do not extend to $T \to 0$\cite{ledwithNonlocal2024,ledwithExotic2025}. The reason is that some hopping sequences which return to the same site are enhanced by powers of $1/T$\cite{vituri2026controlled,nosov2026controlledexpansioncorrelatedelectrons}. These sequences correspond to exchange interactions between fluctuating moments, which eventually lead to the KIVC ordered state at low temperature $T \ll Us^2$. Some insight can be gained into this state by inserting an external field $h$, which plays the role of a self consistent exchange field $h \approx Us^2$ polarizing the electrons into the KIVC state. Then, for $T \ll h$ the trion has vanishing spectral weight and exchange interactions lead to a gapped KIVC spectrum, whereas for $T \gg h$ we recover the gapless electron-trion Dirac cone\cite{ledwithNonlocal2024,ledwithExotic2025}. 

The vanishing spectral weight of the trion at $\bk = 0$ in the ordered state can be understood by noting that, when acting on a uniform background of locally polarized states, the factors $\delta n_\bR$ have the same sign for every site. All Pauli blocking factors are also the same, such that the action of $F_{\bk = \Gamma}$ is proporortional to $\sum_\bR c_\bR$, which vanishes in a topological band. For $Us^2 \gg T > 0$, the correlation length of the KIVC order is long, $> 1/s$, such that this cancellation happens locally almost everywhere.

\putbib[bibtex]
\end{bibunit}

\end{document}